\documentclass[letterpaper,twocolumn,10pt]{article}
\usepackage{usenix}

\usepackage{tikz}
\usepackage{amsmath}
\usepackage{enumitem}
\usepackage{float}

\begin{document}
%-------------------------------------------------------------------------------

%don't want date printed
\date{}

\title{\Large \bf From Speech to Data: Unraveling Google's Use of Voice Data for User Profiling}

%for single author (just remove % characters)
\author{
{\rm Xinhang Ma}\\
Washington University in St. Louis
\and
{\rm Sirui Chen}\\
Washington University in St. Louis
% copy the following lines to add more authors
% \and
% {\rm Name}\\
%Name Institution
} % end author

\maketitle

%-------------------------------------------------------------------------------
\begin{abstract}\label{abstract}
%-------------------------------------------------------------------------------
Smart home voice assistants enable users to conveniently interact with IoT devices and perform Internet searches; however, they also collect the voice input that can carry sensitive personal information about users. Previous papers investigated how information inferred from the contents of users' voice commands are shared or leaked for tracking and advertising purposes. In this paper, we systematically evaluate how voice itself is used for user profiling in the Google ecosystem. To do so, we simulate various user personas by engaging with specific categories of websites. We then use \textit{neutral voice commands}, which we define as voice commands that neither reveal personal interests nor require Google smart speakers to use the search APIs, to interact with these speakers. We also explore the effects of the non-neutral voice commands for user profiling. Notably, we employ voices that typically would not match the predefined personas.%
\footnote{for example, for a profile that Google has inferred to be female, we typically use a male voice to interact with the smart speaker, unless noted otherwise.} We then iteratively improve our experiments based on observations of profile changes to better simulate real-world user interactions with smart speakers. We find that Google uses these voice recordings for user profiling, and in some cases, up to 5 out of the 8 categories reported by Google for customizing advertisements are altered following the collection of the voice commands. 

\end{abstract}

%-------------------------------------------------------------------------------
\section{Introduction}\label{sec:1_intro}

The human voice carries a remarkable amount of personal information, encoding nuances that go beyond mere words. Attributes like tone, pitch, and rhythm can subtly communicate a speaker’s emotions, intentions, and identity~\cite{bonvillain2019language}. While the intricate details embedded in speech might not be discerned by an average listener, they are readily accessible to sophisticated automated voice processing technologies. These technologies have the capability to perform detailed voice profiling, extracting sensitive personal attributes from spoken words. Such attributes can include an individual's age, physiological characteristics, health and medical conditions, personal identity, levels of intoxication, emotional states, stress levels, and even the likelihood of truthfulness~\cite{singh2019profiling, kroger2020privacy, mehta2020recent}. This extensive capability of voice analysis, especially when leveraged by today's advanced algorithms, raises concerns about privacy and data security in the era of smart devices. Our research aims to explore this domain within the context of the Google ecosystem. Specifically, we investigate whether voice interactions with Google smart speakers are used not just for facilitating user-device interactions, but also for constructing detailed user profiles. This study delves into how voice data, even when commands are neutral and non-revealing, might still contribute to the digital portrait Google creates for each user.

The proliferation of Internet of Things (IoT) devices has marked a significant shift in consumer technology, with smart home devices becoming increasingly prevalent~\cite{al2020internet}. As these devices integrate more deeply into everyday life, companies such as Google and Amazon are exploring the potential of voice profiling. Research in this domain indicates a growing interest in utilizing voice data not merely for functional purposes but also for strategic marketing and personalized advertising~\cite{turow2021voice}. For example, to monetize voice data, Amazon has a patent for using voice data to infer the physical and emotional states of users for targeted advertising purposes~\cite{jin2018voice}.

While smart home devices like smart speakers are becoming increasingly popular, users often lack control and insight into how their interactions with these devices are being used. Recent studies reveal that smart speaker platforms, along with their third-party apps, not only collect interaction data but also process it to infer user interests, subsequently using these inferences for targeted advertising~\cite{iqbal2022your}. Building upon these insights, our paper investigates the deeper implications of voice data usage, specifically examining its role in user profiling and targeted advertising in the Google ecosystem.

Previous studies have highlighted that smart speakers, which are often perceived as black-box devices, not only raise privacy concerns for users due to their opaque nature~\cite{lau2018alexa} but also pose difficulties for researchers seeking to understand the specifics of data collection and usage~\cite{iqbal2022your}. In light of this challenge, our research specifically focuses on the Google ecosystem. We posited that the Google My Ad Center~\cite{googleadcenter} could serve as a direct reflection of Google's inferences about its users. This approach allows us to circumvent some of the opacity inherent in these devices, and through controlled experiments, allows us to provide a unique window into how user interactions, particularly voice commands, are processed and utilized for profiling purposes.

To conduct our experiments, we first build user profiles that reflected various interest personas. This process involves the use of an automated bot that we programmed to browse the internet via a Chrome browser. The bot engages with content that corresponded with the interests of each predefined persona, thus creating distinct digital profiles. After this, we introduce voice data into the mix. This step involves interactions with Google smart speakers with carefully-crafted neutral voice commands. These commands are designed to contain no content that could reveal the persona's specific interests, focusing instead on the vocal characteristics themselves. This method is crucial in assessing the influence of voice data on user profiling, particularly in how Google's algorithms process and respond to these voice characteristics, separate from the actual content of the commands.

To determine if voice data is indeed utilized for user profiling, our primary tool of analysis is the Google My Ad Center. This platform provides a direct insight into how Google's algorithms might be interpreting and utilizing voice interactions for tailoring user profiles. Furthermore, after conducting preliminary tests, we engage in follow-up experiments. These are driven by observations and patterns that emerged from our initial findings. These subsequent experiments allow us to investigate into specific aspects of voice data usage, offering a more nuanced understanding of the profiling process in the Google ecosystem. By iteratively refining our approach based on experimental results, we discover and document how Google might be using voice data to profile its users for tracking and advertising purposes. 

Through the course of our experiments, we observed notable changes in the way Google profiles its users based on their voice data interactions. Specifically, in several instances, we documented shifts in up to 5 out of the 8 categories used by Google for customizing advertisements. These changes occurred subsequent to the introduction of voice data through our controlled interactions with Google smart speakers. To facilitate further research and transparency, we have released our code at \url{https://github.com/xhOwenMa/voice_data_google}, which contains thorough guidelines and examples.
\section{Background}\label{sec:2_background}

In this section, we provide the background to contextualize our research within the broader spectrum of relevant topics. First we introduce the Google smart speaker platform. We then explore the concept of voice profiling, examining how voice data can be processed and utilized for various purposes. Further, our discussion extends to online tracking and ad targeting mechanisms. This background serves to lay the groundwork for understanding the complexities and implications of our findings in the realm of voice data usage by Google.

\textbf{Google smart speakers:} In this paper, we focus on Google's smart speaker platform, a prominent player in the market with a significant user base~\cite{globalspeakermarketshare}. Google's smart speakers, now known under the Google Nest brands~\cite{googlenestspeaker}, are integrated with Google Assistant~\cite{googleassistant}, a voice-activated helper designed to respond to a variety of voice commands. As the smart speaker market size continues to grow, it raises important questions about the privacy implications of voice data collection and usage. 

\textbf{Voice profiling} harnesses the unique characteristics of speech to draw inferences about individuals' identities and states. The ability to estimate characteristics like age from a voice has been within reach for decades~\cite{ptacek1966age}. Recent explorations in the field have expanded these capabilities, investigating the potential to deduce even facial features~\cite{wen2019face, oh2019speech2face} and emotional states~\cite{mcgilloway2000approaching} from vocal inputs. The application of such analysis varies widely, providing innovative ways to interact with technology and expanding service accessibility, for example, speech-based systems have been developed to aid community health workers in rural areas with limited literacy, allowing them to access vital health information~\cite{sherwani2007healthline}.

Building on the foundational capabilities of voice profiling, numerous studies have demonstrated its effectiveness in speaker verification and recognition~\cite{dehak2010front, snyder2018x, chung2018voxceleb2, nagrani2017voxceleb}. This technological proficiency is actively employed in smart home speakers to deliver customized user experiences~\cite{googlevoicematch, alexavoiceid}. However, it also raises privacy concerns. In the intimate setting of one's home, a smart speaker that can discern and interpret various nuances of speech inherently has access to a significant amount of personal information. While this facilitates convenience and personalization, it also poses a potential risk for privacy breaches. There is a growing need to address how these devices handle sensitive voice data as studies have found that both direct users and those indirectly affected express concerns around the use of their data~\cite{malkin2019privacy, meng2021owning}.

\textbf{Online tracking \& Ad targeting:} given that the focus of our research is on how voice data is processed for advertisement customization in the Google ecosystem, online tracking stands as an inherently connected field of study. The practices of online tracking have long been scrutinized for their implications on privacy~\cite{englehardt2016online, iqbal2021fingerprinting, west2019data}, as they collect a vast array of user data to deliver targeted advertising. In the realm of smart home speakers, voice data represents a novel and rich source of information that could be leveraged in similar ways~\cite{iqbal2022your}. Our investigation centers on Google's handling of voice interactions, examining the extent to which these exchanges are used to customize advertisements to the individual. 
\section{Methodology}\label{sec:3_method}

In this section, we describe our methodology, detailing the experimental approach we adopted to test and document the ways in which Google utilizes voice data to profile its users. Our method consists of three steps: we first simulate user personas with distinct digital behaviors to establish baseline profiles. These personas are then used to interact with Google's smart speakers through neutral voice commands that are crafted to avoid revealing specific user interests. Finally, We apply an iterative approach, refining our tests based on continuous observations to capture the nuances of Google's use of voice data.

\subsection{Building Baseline Profiles}\label{subsec:3.1_baseline_profiles}

To establish the baseline for our experiments, the initial step involved creating a series of user personas. These personas were designed to reflect a diverse range of interests and behaviors, simulating real-world users with different browsing habits and interaction patterns. 
\begin{figure}
    \centering
    \includegraphics[scale=0.64]{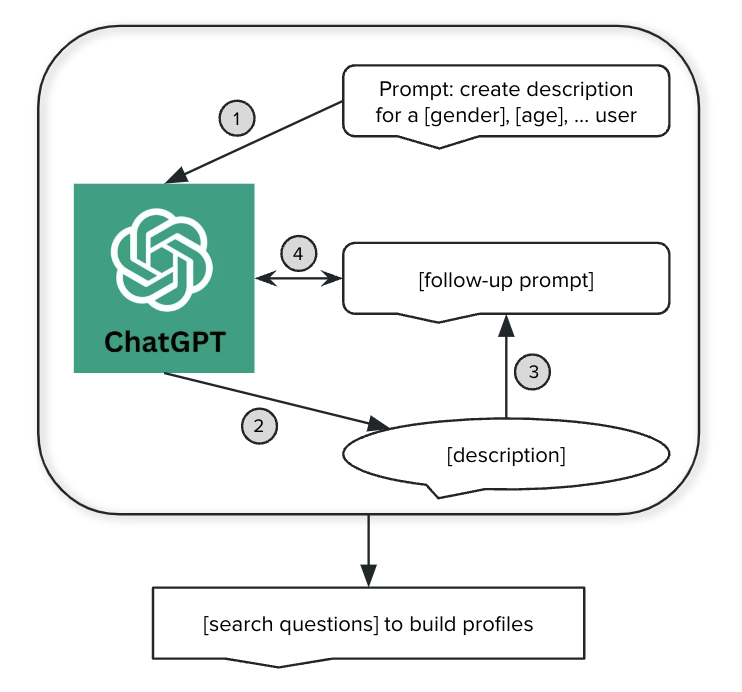}
    \caption{Overview of how to build baseline profiles: step 4 is repeated until we are confident that the resulting search questions effectively represent the intended characteristics of each persona.}
    \label{fig:build_baseline}
\end{figure}

Figure~\ref{fig:build_baseline} presents the overview of our approach for creating baseline profiles. The process began with prompting ChatGPT with a high-level description for each user persona, only specifying attributes such as gender and age. Upon receiving the initial response, we devised subsequent prompts to narrow down the focus, selecting between five to seven categories from ChatGPT's response. For each category, we then asked ChatGPT to produce a list of 10 related search questions. These questions, after undergoing manual reviews to ensure their relevance and accuracy, were given to our automated browsing bot. The bot executed these queries across the internet for ten iterations, effectively simulating the digital activity of our user personas and thereby constructing their baseline profiles. 

Our automated browsing bot is designed to mimic human browsing behavior. Once initiated, it opens a Chrome browser session logged into a profile corresponding to one of our simulated personas. The bot is programmed to input the search queries into the browser, just as a user might. It then navigates the search results, clicking on links to visit websites, scrolling through web pages to simulate engagement, and interacting with common web prompts such as cookie consent dialogs. These interactions are crucial for creating a realistic digital footprint and circumventing bot detection. To validate the effectiveness of our bot, we compared the completeness of the profiles it developed with those created by OpenWPM~\cite{englehardt2016online}, and the results confirmed its functionality. Compared to the OpenWPM tool, which serves broader objectives and therefore is much more complicated to set up and configure, our bot demonstrated that it could operate with a more lightweight and user-friendly approach for studies with goals similar to ours. 

\subsection{Interacting with the Smart Speakers}\label{subsec:3_2_voice_interactions}

Upon establishing the baseline profiles, we proceeded to engage with Google smart speakers through the use of neutral voice commands. Our experiments were conducted using the Google Nest Mini 2nd Generation. It can be assumed that our findings are applicable to the broader range of Google smart assistant devices, as they share the common thread of being powered by Google Assistant ~\cite{googleassistant}, which suggests that Google processes voice data consistently across its device spectrum. We leave the actual experiments on alternative devices to future works.

To introduce voice data effectively, we first established a set of \textit{neutral voice commands}. These commands are designed to be non-revealing of the user's personal interests and do not trigger the smart speakers' search APIs. To create the neutral voice commands, we adopted a similar method as in Section~\ref{subsec:3.1_baseline_profiles}. The key variation is that we utilized in-context few-shot learning~\cite{brown2020language, min2022rethinking} techniques to prompt ChatGPT with three exemplars which we manually validated to satisfy our definition of neutral voice commands as the foundation. This approach enabled us to generate a diverse array of commands. Then we used OpenAI's text-to-speech (TTS) APIs~\cite{openaitts} to convert these commands into speech. We selected the Echo voice option for male voice and the Shimmer voice option for female voice. 

We developed eight to ten pairs of these neutral voice commands for each baseline profile, each pair consisting of an action and a corresponding command to reverse it. For instance, one such pair includes the command \textit{Hey Google, set a timer for 20 minutes}, along with its counterpart to cancel the timer, \textit{Hey Google, cancel the timer}. Then we used these commands in our interactions with the smart speaker. Crucially, we used a male voice for profiles inferred as female, and a female voice for those inferred as male. Here we aim to answer the research question:
\begin{description}
    \item [RQ1:] \textit{Is voice data%
    \footnote{aside from gender, we also tested age and language in the first round of testing} used for user profiling in Google ecosystem?}
\end{description} 
Additionally, it is noteworthy that for all experiments, once we began introducing voice data through these commands, we stopped all web browsing activities with the profiles. This approach ensured that any changes observed in the Google My Ad Center could be attributed solely to the voice interactions.%
\footnote{According to Google Nest's privacy policies~\cite{googlevoicedatac}, data from interactions with these devices can influence the ads users receive, effectively modifying profiles in My Ad Center.} In Section~\ref{subsec:3_3_refining_experiments}, we detail our follow-up experiments which include incorporating additional nuances to more accurately mimic real-world user interactions with smart speakers and other relevant aspects.

\subsection{Refining Experiments}\label{subsec:3_3_refining_experiments}

Following the profile changes in the initial round of testing, we took steps to refine and expand the scope of our experiment based on these initial observations. In this section, we outline the key observations that motivated us and how they shaped the corresponding research questions for our subsequent experiments:
\begin{description}
    \item [RQ2:] \textit{How much can Google infer about a user from one neutral voice command?}
\end{description}
This follow-up experiment was directly inspired by the profile changes we observed in the first phase of our experiments. Initially, all of our experiments were based on the hypothesis that profile changes were the cumulative effect of multiple voice interactions. However, in order to precisely understand Google's handling of voice data, we created a fresh profile and limited our interaction with the smart speaker to only one specific neutral voice command — the same command that immediately preceded the profile change in the initial phase. For example, if a category change was noted after using the command \textit{Hey Google, remind me to take out the trash tonight}, we would replicate this command with the smart speaker on a new profile absent of prior web browsing or voice data. Specifically, a fresh profile only contains age and language recorded during registration and gender is not shown
in My Ad Center because we chose "prefer not to say" during account creation. Our aim was to isolate the impact of a potentially influential voice command and determine how much Google could infer about a user from this interaction alone.

\begin{description}
    \item [RQ3:] \textit{For profile that already contain voice data, will different voice characteristics alter the existing profile?} 
\end{description}
This reflects a common real-world situation where smart speakers, typically placed in shared spaces, might collect voice data from both the device's owners and their roommates or visitors. Furthermore, this opens up possibilities for investigating methods to potentially counteract profiling practices by interacting with smart speakers using artificially generated voices.

To conduct these experiments, voice data was introduced at the stage of baseline profile creation, with the voice gender matching the profile's inferred identity. Once these baseline profiles were established, we then interacted with the smart speakers using a voice gender different from that initially used. This approach allowed us to assess the impact of introducing a contrasting voice characteristic on the already formed user profiles.

\begin{description}
    \item [RQ4:] \textit{How will non-neutral voice commands be used for profiling?}
\end{description}
This direction is particularly relevant given that in real-world scenarios, users often interact with their smart speakers in ways that could disclose personal preferences. To conduct this experiment, we applied non-neutral voice commands across two types of baseline profiles. The first type, \textit{web profiles}, were established through web browsing only. The second type, \textit{voice profiles}, included both web browsing data and previous voice interactions. This approach allowed us to examine the influence of non-neutral voice commands across differently constructed profiles. The experiments were conducted in a manner consistent with our established methodology, ensuring comparability and continuity in our investigative process.

In Appendix~\ref{appendix_B:examples}, we provide the prompts and resulting search questions and voice commands used for building baseline profiles as well as the subsequent experiments. They are also included in our Github repository.
\section{Experimentation \& Results}\label{sec:4_results}

In this section, we assess the implications of our experiments on user profiling as perceived through Google My Ad Center~\cite{googleadcenter}. First, we outline the specific categories within Google My Ad Center that were the focus of our study, providing a framework for understanding the changes observed. Then, we unpack the results obtained from our various experimental setups, including the initial round of testing with neutral voice commands and the subsequent follow-up experiments.  This section aims to provide a clear and thorough presentation of our findings, highlighting the nuanced ways in which voice data influences user profiles in the Google ecosystem.

\subsection{Categories Used for Evaluations}\label{subsec:4_1_adcenter_categories}

\begin{table}[]
    \centering
    \includegraphics[scale=0.6]{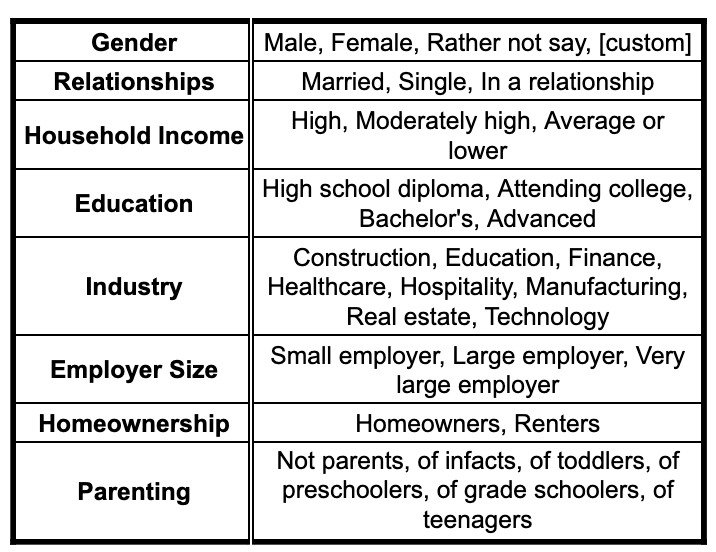}
    \caption{Categories from Google My Ad Center that are used for evaluations. For Gender, we did not consider any `custom' genders due to difficulties of testings.}
    \label{tab:adcenter_categories}
\end{table}
Table~\ref{tab:adcenter_categories} shows the categories used to evaluate changes in user profiles within Google My Ad Center. These categories were selected to cover the demographic and socio-economic spectrums relevant to user profiling for targeted advertising. Each category provided a range of options that allowed for granular analysis of the inferred profile changes resulting from interactions with the smart speaker.

\subsection{Google Uses Voice Data for Profiling}\label{subsec:4_2_neutral_voice_after_baseline}

To answer \textbf{RQ1:} \textit{Is voice data used for user profiling in Google ecosystem?} We utilized neutral voice commands spoken in a gender opposite to the one inferred in the baseline profile to observe potential shifts in profile categorization. We found that, even with the employment of neutral voice commands, Google's inferred user profiles did indeed change. This suggests that voice data, irrespective of the content of the commands, contributes to the profiling criteria within the Google ecosystem. This initial set of results serves as an indication that Google's algorithms consider voice characteristics as a factor in user profiling. 
\begin{table}
    \centering
    \includegraphics[scale=0.62]{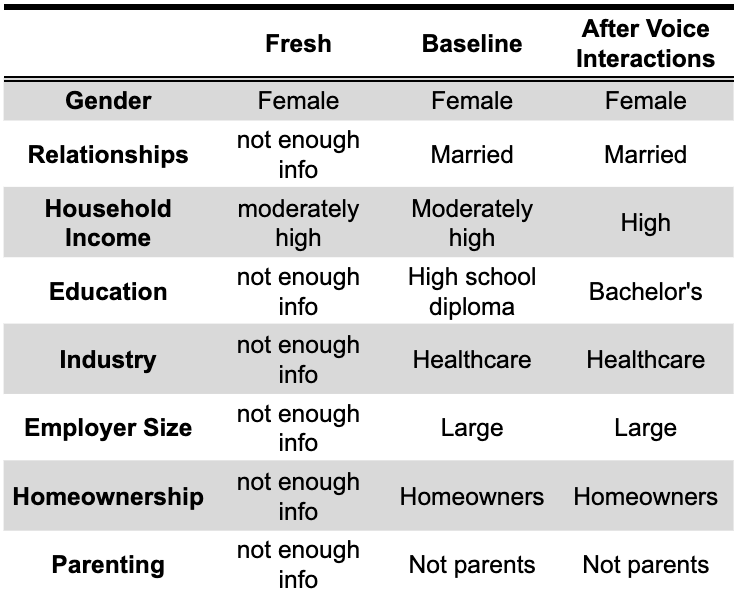}
    \caption{Profile changes across different interaction phases (neutral voice commands after baseline profile is built): `Fresh' indicates the initial inferred attributes without any interaction; `Baseline' reflects the attributes after simulating web browsing behavior; `After Voice Interactions' shows the attributes following the introduction of neutral voice commands using the opposite gender's voice.}
    \label{tab:4_2_profile_change_after_neutral_voice_commands}
\end{table}

Table~\ref{tab:4_2_profile_change_after_neutral_voice_commands} presents findings from one of our user profiles during the first phase of experimentation, demonstrating the evolution of Google's inferred attributes through different stages of interaction. Initially, the profile was considered `Fresh' with minimal information available. To establish a `Baseline', our automated browsing bot conducted a series of web searches, repeating 60 search queries across six categories such as interests, hobbies, family, and lifestyle, iterating them 10 times to simulate a realistic user browsing pattern. After establishing this baseline profile, we introduced neutral voice commands, executing $10$  pairs of neutral voice commands using a male voice. An important consideration in interpreting the results is the nature of the activity history created by voice commands. Typically, these are categorized as Assistant AI activities, but they can also trigger web searches, podcast interactions, and other types depending on the command's requirements for response. If any neutral voice command created activities other than Assistant AI activities, we deleted that activity and also deleted that voice command from our experiment. 

From Table~\ref{tab:4_2_profile_change_after_neutral_voice_commands}, we can observe that, post interaction, notable changes were observed in several categories such as `Household Income' and `Education', indicating that Google's profiling system had updated the user attributes based solely on voice data. Notably, these alterations happened from a relatively small number of voice commands (approximately $50$), especially in comparison to the extensive web searches used to construct the baseline profile.  Such findings highlight the influence of voice interactions on user profiling within the Google ecosystem, even when such interactions are designed to be neutral and non-revealing of personal preferences. 

However, the fact that the gender category remained unchanged, despite our deliberate use of a male voice for a profile inferred as female, is unexpected. Our hypothesis is that the number of voice interactions was insufficient to influence Google's established gender inference. Another consideration is the sophistication of Google's voice recognition technology, which might be able to distinguish between genuine user voice alterations and those that are artificially introduced or insufficiently varied. Further research would be necessary to explore these possibilities and to determine the precise thresholds and mechanisms Google employs for updating user gender attributes based on voice data.

\subsection{Single Neutral Voice Command}\label{subsec:4.4_single_voice_command}
To answer \textbf{RQ2:} \textit{How much can Google infer about a user from one neutral voice command?} We created two fresh profile and selected the two neutral voice commands that directly led to profile changes in our initial experiments. Specifically, these voice commands are \textit{Hey Google, what's the capital of Brazil?} and \textit{Hey Google, Remind me to call the dentist at 11 AM.}

In the previous experiment, profile changes were observed after approximately 50 voice interactions. The experiment scheme here is to isolate potentially influential voice commands by checking if categories shown in My Ad Center change after feeding such commands to a completely fresh account with no history of data collection. 

Our experiment revealed a cumulative effect of voice commands on user profiles. While we spaced out the voice commands, a significant and sudden update in the My Ad Center profile occurred after a certain number of interactions. Interestingly, some of these updates seemed unrelated to the commands themselves. For example, updates to `Employment' categories occurred following the command: \textit{Hey Google, what's the capital of Brazil?} 

This cumulative effect in profile updating, rather than changes being triggered by single voice commands, was replicated consistently, with profiles beginning to show updates in My Ad Center after accumulating around 50 voice history entries. This finding indicates that Google's profiling algorithm may require a threshold of voice data interactions before reflecting changes in user profiles. However, it is also worth considering that this high threshold could be related to our use of neutral voice commands. In Section~\ref{subsec:non_neutral}, we discuss non-neutral voice commands, which are inherently more revealing, and explore their potential impact on user profiling.

\subsection{Contrasting Voice Characteristics has Minimal Impact on Established User Profiles}\label{subsec:4_4_contrasting_voice}

To answer \textbf{RQ3:} \textit{For profile that already contain voice data, will different voice characteristics alter the existing profile?} We augmented our methodology for establishing baseline profiles. We incorporated voice interactions that were congruent with the simulated personas, mirroring the actions a real user would likely take. For example, for a profile representing a high school mom persona, we included voice commands like \textit{"Hey Google, show me easy-to-prepare school lunch recipes,"} to enrich the profile with relevant voice data. Once the baseline profile was built with these tailored voice interactions, we introduced neutral voice commands in a male voice. This allowed us to observe if and how the profile would adjust to new voice characteristics that differed from the established persona. The goal was to assess the adaptability of Google's user profiling to changes in voice data that could occur in shared environments, such as a household with multiple residents.

The experimental results indicated that the user profile remained unchanged after approximately $50$ neutral voice command interactions. This number is notably small in comparison to the volume of web browsing and it is even smaller than the voice interactions conducted during the baseline profile building.%
\footnote{we had 6 different voice commands and they were played to the smart speaker 10 times each during baseline profile building.} This suggests that a higher frequency of voice interactions may be required to influence the profile. However, considering that we had already conducted 50 interactions with the smart speaker (the same quantity as in our experiments detailed in Section~\ref{subsec:4_2_neutral_voice_after_baseline}), we concluded that further experiments with neutral voice commands would be unlikely to yield different outcomes on baseline profiles that are well-established. Consequently, we decided to broaden the scope of our voice commands to include non-neutral variants, aiming to capture a wider spectrum of user interactions with the smart speaker.

\subsection{Non-Neutral Voice Commands}\label{subsec:non_neutral}

In conducting the experiments on non-neutral voice commands to answer \textbf{RQ4:} \textit{How will non-neutral voice commands be used for profiling?} We applied these commands to both the web profiles and voice profiles for testing. Throughout this section, we will focus on a baseline profile initially identified by Google as female. Parallel experiments were also conducted on a baseline profile simulating a male persona. For more details on these parallel experiments, readers are directed to our GitHub repository.

To make sure that non-neutral voice commands are actually influential, we conducted tests on a freshly created profile, results after only $10$ voice interactions are detailed in Table~\ref{tab:non-neutral-on-fresh}. Specifically, we aimed to construct a profile portraying a male individual working in the finance field with a high income and marital status of married, but interacted with the smart speaker using a female voice. Remarkably, with as few as 10 voice interactions, there were significant changes in all of the categories within Google My Ad Center: the gender was inferred as female, aligning with the voice used during interactions, while accurately capturing the high-income bracket and finance employment from the content of the voice commands. This outcome highlights the dual influence in the non-neutral voice commands – the voice tone impacting the gender categorization and the command content shaping professional and income-related aspects of the profile. Another observation was that activity history updates associated with Assistant AI tended to lag behind those of standard web search activities. This difference became particularly evident here because non-neutral voice commands often logged searches and other activities, leading to quicker profile updates. In contrast, the slower update rate for histories associated with neutral voice commands via Assistant AI partly explains the necessity for a larger volume of interactions to observe changes in the user profile.

\begin{table}[]
    \centering
    \includegraphics[scale=0.45]{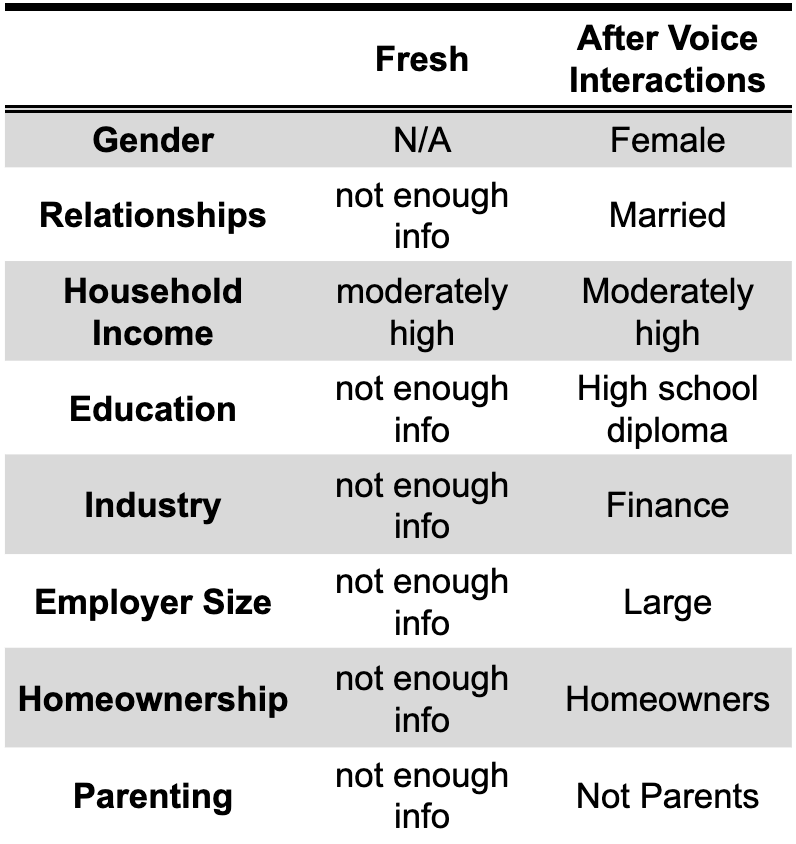}
    \caption{Profile changes: after $10$ non-neutral voice command interactions on a fresh profile}
    \label{tab:non-neutral-on-fresh}
\end{table}

Then we tested the effects of non-neutral voice commands on web profiles. We observed significant changes in the user profiles after approximately 50 voice interactions. This pattern of change aligns with what we observed using neutral voice commands, but the alterations in profile categories were more significant with non-neutral commands. Our strategy was to deliberately contrast the established profile attributes with our voice commands. For example, for a profile initially categorized as female with high income, we would employ a male voice for the commands and crafted them to suggest a lower income bracket. As shown in Table~\ref{tab:non-neutral-web-profile}, there are noticeable changes in `Relationship', `Household Income', and `Industry' categories. These shifts in profile attributes were not only more substantial than those observed with neutral commands but also more directly correlated with the specific content of the non-neutral voice interactions. After an additional 100 voice interactions, we observed changes in categories `Education' and `Employer Size'. This suggests that the quantity of voice interactions plays a critical role in the extent and depth of profile changes. The cumulative effect of these interactions likely contributed to the broader shifts observed in the user profile attributes.

\begin{table}[]
    \centering
    \includegraphics[scale=0.65]{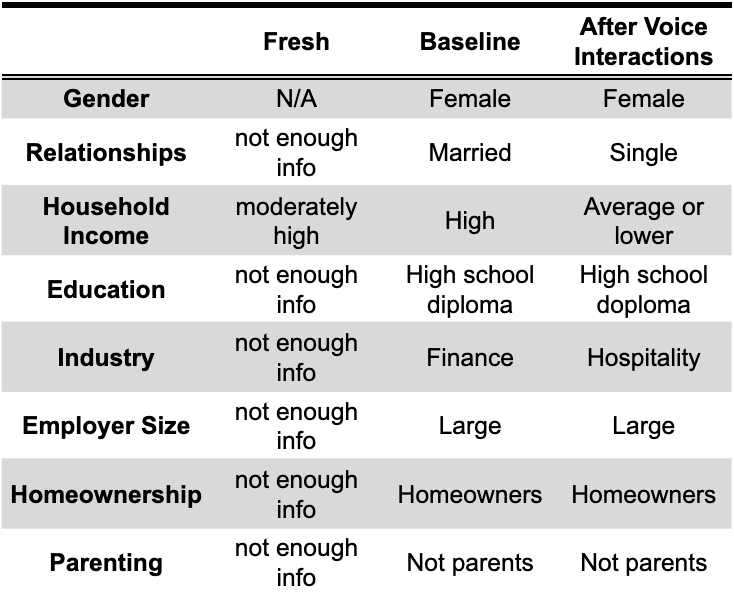}
    \caption{Profile changes: $50$ non-neutral voice command interactions on a web profile}
    \label{tab:non-neutral-web-profile}
\end{table}

When analyzing the impact of non-neutral voice commands on voice profiles, our observations indicated fewer changes compared to the web profiles. This suggests that the existing voice data in these profiles may have established a stronger baseline. Notably, there were instances where the interactions did not get reflected in the Google Activity history. This phenomenon raises questions about the processing and weighting of voice data within Google's algorithms, especially when the profile already contains a significant amount of voice interaction history. It may also indicate a level of sophistication in the profiling system, where it recognizes and maintains consistent user patterns over contradictory new data. We leave the exploration of thresholds for impactful voice data on existing voice-enriched profiles to future studies.

\section{Limitation \& Discussion}\label{sec:limitation&discussion}

In this section, we contextualize our results within the wider scope of user privacy, profiling practices, and the landscape of smart technologies. We first discuss the constraints under which our experiments were conducted, then we explore the implications of our study for users, industry practices, and future research. 

\subsection{Limitation: Profile and Device Variability}\label{subsec:limited_profiles_&_devices}

Our experiment was conducted with a finite set of profiles and only three Google smart speakers. While these were sufficient to observe and document changes in user profiling based on voice data, this limited sample size means that our results may not be statistical significant across the broader spectrum of users and devices. Therefore, our findings should be interpreted conservatively when considering their applicability to the entire population of smart speaker users. This limitation points to the need for further research involving a larger and more varied sample to corroborate and expand upon our findings, ensuring a more comprehensive understanding of voice data usage in user profiling. 

Additionally, our experiments were also constrained by scope of our interactions, both in terms of web browsing and voice commands. The profiles we constructed did not encompass the full range of interactions of a user who is more engaged with digital content. Future studies could benefit from exploring more involved profiles that include a diverse range of digital interactions such as extensive web browsing, YouTube viewing habits and preferences, and a wider array of voice interactions. Profiles incorporating these aspects would present a more comprehensive representation of an engaged user’s behavior, potentially revealing deeper insights into how Google's profiling algorithms synthesize multifaceted user interactions. 

\subsection{Discussion: Why is `Gender' Unchanged}

A notable observation across our experiments is that the gender category within the user profiles remained unchanged, despite our deliberate use of voice commands in a contrasting gender. This persistence raises questions about the specific criteria Google's algorithms use to infer and update gender. One possible explanation is that Google may assign a higher weight to initial data points used to establish gender or may require a consistent and prolonged pattern of contrary indicators to update this particular attribute. This observation highlights the complexity and potential conservatism in how certain fundamental user attributes are managed and updated within profiling systems. Further research to understand the mechanisms and thresholds that govern such modifications in user profiling systems is important, especially in light of evolving discussions about gender identity and expression in digital spaces.

\subsection{Implications: Ethics and Privacy of Voice Data Profiling}\label{subsec:ethics_privacy_implication}

The findings of our study open up discussions on the ethical and privacy implications of voice data collection and user profiling. Our research highlights the sensitivity of these devices to voice interactions and the potential for significant user profiling based on limited voice data. This raises concerns about user consent and awareness, especially given that previous research had indicated that smart speakers may record audios from their surroundings even when they should not~\cite{dubois2020speakers}. Hence, our findings point to the need for more robust and transparent privacy policies from tech companies, ensuring users are informed and have greater control over their data. Future research should consider the balance between technological advancements and the protection of individual privacy rights in light of evolving user data usage dynamics.

\subsection{Implications: Voice Interactions as A Tool for Obfuscation}\label{subsec:voice_as_obfuscation}

An interesting implication of our findings is the potential use of voice interactions as a tool for obfuscation, deliberately distorting user profiles to enhance privacy. The sensitivity of Google's profiling system to voice data suggests that intentionally varied or inconsistent voice interactions through non-neutral voice commands could potentially introduce ambiguity into the user profiling process. For instance, an individual might utilize specific voice commands that are inconsistent with their actual personal preferences and information, thereby diminishing the precision of their digital profiles. Future exploration in this area could shed light on the feasibility and effectiveness of such tactics as a means of safeguarding personal privacy in an environment where digital footprints are increasingly leveraged for commercial purposes.
\section{Ethics}\label{sec:5_ethics}

Given the nature of our study, we conscientiously adhered to ethical standards throughout the process. All profiles used for experimentation were created and simulated, ensuring that no real individuals' data or privacy were compromised. Additionally, our automated browsing was conducted on a limited scale and designed to minimize any potential interference or distortion of web traffic patterns.

However, in our experiments involving non-neutral voice commands, there is a potential for these commands to carry traditional gender stereotypes. This was not the intention of our study, but it is an unavoidable aspect of attempting to simulate real-world interactions, which themselves can often be embedded with societal norms and biases. To ensure full transparency and provide context, we have made all the voice commands employed in our experiments publicly available in our GitHub repository. 
\section{Conclusion}\label{sec:6_conclude}

In this paper, we explored how voice data is collected and used for user profiling in Google ecosystem. Our results indicate that both neutral and non-neutral voice commands have the potential to alter user profiles, although at different intensities. 

The use of neutral voice commands revealed the capability of voice data to influence certain aspects of the profile, such as professional and income categories. However, these changes required a substantial number of interactions, suggesting a cumulative effect rather than immediate impact. In contrast, non-neutral voice commands demonstrated a more significant and immediate influence on user profiles, underscoring the sensitivity of Google's profiling algorithms to voice data that is more revealing of personal preferences and characteristics. In conclusion, our study contributes to the growing body of knowledge on digital privacy and user profiling, offering insights into the complex dynamics of voice data within the Google ecosystem. As smart devices become increasingly integrated into daily life, understanding these dynamics becomes crucial for both users and policymakers alike.

%-------------------------------------------------------------------------------
\bibliographystyle{plain}
\bibliography{ref}

\appendix

\section{Examples of ChatGPT Prompts and Answers}\label{appendix_B:examples}

\subsection{Building Baseline Profiles}
\begin{table}[ht]
    \centering
    \begin{tabular}
    {|p{0.95\linewidth}|}
    \hline 
      \textbf{Q: }I am building a Google profile for the following user: [gender], [age], [and maybe a couple more very general description]. Provide me with a general description of this user. \\
      \vspace{1mm}
      \textbf{A: }... you might consider the following general characteristics and interests:
      \setlist{nolistsep}
      \begin{itemize}[noitemsep]
          \item Demographics: ...
          \item Interests and Hobbies: ...
          \item Technology Usage: ...
          \item Family and Lifestyle: ... 
      \end{itemize}
      (truncated) \\
      \vspace{1mm}
      \textbf{Q: }for each category, give me concrete descriptions that will make this profile as real as possible \\
      \vspace{1mm}
        \textbf{A: }... here's a more specific characterization:
        \setlist{nolistsep}
      \begin{itemize}[noitemsep]
          \item Demographics: ...
          \item Interests and Hobbies: ...
          \item Technology Usage: ...
          \item Family and Lifestyle: ... 
      \end{itemize}
      (truncated) \\
      \vspace{1mm}
      \textbf{Q: }based on this profile description, generate 10 possible search questions that are related to this profile per category. \\
      \vspace{1mm}
      \textbf{A: }[ChatGPT proceeds to give $10$ search questions per category, and we use these to build the baseline profile] \\
      \hline
    \end{tabular}
    \caption{Example prompts to create profile descriptions and search questions for web browsing to build the baseline profiles}
    \label{tab:prompts-for-building-baseline}
\end{table}

\subsection{Neutral Voice Commands}

% \hfill
% \vline 
% \hfill

\begin{table}[H]
    \centering
    \begin{tabular}{|p{0.95\linewidth}|}
    \hline
        \textbf{Q: }I am interacting with Google smart speaker, give me 10 neutral voice commands and their corresponding undo commands. Here are a few examples of neutral voice commands: 
        \begin{enumerate}[noitemsep]
            \item 
            \begin{itemize}[noitemsep]
                \item Hey Google, set a timer for 30 minutes for the oven.
                \item Hey Google, cancel the oven timer.
            \end{itemize}
            \item \begin{itemize}[noitemsep]
                \item Hey Google, remind me to take out the trash tonight.
                \item Hey Google, cancel my reminder to take out the trash.
            \end{itemize}
        \end{enumerate}\\
        \vspace{1mm}
        \textbf{A: }... :
        \setlist{nolistsep}
        \begin{enumerate}[noitemsep]
            \item \begin{itemize}[noitemsep]
            \item "Hey Google, remind me to water the plants tomorrow morning."
            \item "Hey Google, cancel my reminder to water the plants."
        \end{itemize}
        \item \begin{itemize}[noitemsep]
            \item "Hey Google, add apples and bananas to my shopping list."
            \item "Hey Google, remove apples and bananas from my shopping list."
        \end{itemize}
        \item ...
        \end{enumerate}
        (truncated)
         \\
        \hline
    \end{tabular}
    \caption{Examples of neutral voice commands}
    \label{tab:my_label}
\end{table}

\subsection{Non-Neutral Voice Commands}

For these commands, we found the most straightforward way to create them is by continuing a previous chat for creating baseline profiles and ask Chat-GPT to generate voice commands that are likely to be used by that user.

\begin{table}[H]
    \centering
    \begin{tabular}{|p{0.95\linewidth}|}
    \hline
        (continue from previous chat for creating profile descriptions and search questions) \\
        \vspace{1mm}
        \textbf{Q: }generate 10 voice commands this user will most likely use to interact with google smart speaker. \\
        \vspace{1mm}
        \textbf{A: }Based on the profile ... :
        \setlist{nolistsep}
        \begin{itemize}[noitemsep]
            \item "Hey Google, add a family meeting to my calendar for next Saturday at 3 PM."
            \item "Ok Google, find a recipe for a healthy vegetarian lasagna."
            \item "Hey Google, set a reminder for my yoga class tomorrow at 6 AM."
            \item ...
        \end{itemize}
        (truncated) \\
        \hline
    \end{tabular}
    \caption{Examples of non-neutral voice commands}
    \label{tab:my_label}
\end{table}

%%%%%%%%%%%%%%%%%%%%%%%%%%%%%%%%%%%%%%%%%%%%%%%%%%%%%%%%%%%%%%%%%%%%%%%%%%%%%%%%
\end{document}